\documentclass[12pt,a4paper]{article}
\usepackage{graphicx}
\usepackage{amsmath}
\usepackage{amssymb}
\usepackage{enumerate}
\usepackage{natbib}

\newcommand{\eref}[1]{(\ref{#1})}

\newcommand{\E}{\textrm{E}}
\newcommand{\dd}{\textrm{d}}
\newcommand{\doo}{\textrm{do}}
\newcommand{\N}{\textrm{N}}
\newcommand{\Jyvaskyla}{Jyv{\"a}skyl{\"a}}

\newcommand\independent{\protect\mathpalette{\protect\independenT}{\perp}}
\def\independenT#1#2{\mathrel{\rlap{$#1#2$}\mkern2mu{#1#2}}}

\begin{document}
\title{Estimating complex causal effects from incomplete observational data}
\author{Juha Karvanen\\
Department of Mathematics and Statistics,\\
University of \Jyvaskyla,\\
\Jyvaskyla, Finland\\
juha.t.karvanen@jyu.fi}

\maketitle

\section*{Abstract}
Despite the major advances taken in causal modeling, causality is still an unfamiliar topic for many statisticians. In this paper, it is demonstrated from the beginning to the end how causal effects can be estimated from observational data assuming that the causal structure is known. To make the problem more challenging, the causal effects are highly nonlinear and the data are missing at random. The tools used in the estimation include causal models with design, causal calculus, multiple imputation and generalized additive models. The main message is that a trained statistician can estimate causal effects by judiciously combining existing tools.


~\\
\noindent Keywords: causal estimation, data analysis, missing data, nonlinearity, structural equation model

\section{Introduction}

During the past two  decades major advances have been taken in causal modeling. Despite the progress, many statisticians still hesitate to talk about causality especially when only observational data are available. Caution with causality is always advisable but this should not lead to avoidance of the most important statistical problems of the society, that is, predicting the consequences of actions, interventions and policy decisions.

The estimation of causal effects from observational data requires external information on the causal relationships. The dependencies seen in observational data may be a result of confounding instead of causality. Therefore, the observation data alone are not sufficient for the estimation of causal effects but must be accompanied with the causal assumptions. It suffices to know the causal relationships qualitatively, i.e. to know whether $X$ affects $Y$, $Y$ affects $X$ or $X$ and $Y$ do not have an effect to each other. Obtaining this kind of information may be very difficult in some situations but can also be straightforward if  the directions of the causal relationships can be easily concluded e.g. from the laws of physics or from the temporal order of the variables. 

Structural causal models offer a mathematically well-defined concept to convey causal assumptions and causal calculus provides a systematic way to express the causal effects in the terms of the observational probabilities \citep{Pearl:1995a,Pearl:book}. The benefits of this framework are numerous: First, the philosophically entangled concept of causality has a clear and practically useful  definition. Second, in order to tell whether a causal effect can be identified in general it is sufficient to specify the causal model non-parametrically, i.e. specify only the causal structure of the variables. Third, the completeness of causal calculus has been proved \citep{Pearlsiscomplete,Bareinboim:zidentifiability} and algorithms for the identification of the causal effects have been derived \citep{tian2002general,shpitser2006identificationconditional,shpitser2006identificationjoint,Bareinboim:zidentifiability}.
Fourth, the framework can be extended to deal with study design, selection bias and missing data \citep{dagdesign,Bareinboim:controllingselectionbias,Mohan:missingdata}.

 
Estimation of nonlinear causal effects from the data has in general received only a little attention in the research on structural causal models. Examples of causal estimation in the literature often use linear Gaussian models or consider causal effects of binary treatments.  With linear Gaussian models causal effects can be often directly identified in closed-form using path analysis as recently reminded by \citet{pearl2013linear}. However, in practical situations, the assumption of the linearity of all causal effects is often too restrictive. G-methods \citep{hernanrobisn:book}
and targeted maximum likelihood estimation (TMLE) \citep{vanderlaan2006targeted,vanderlaan2011targeted} are flexible frameworks for causal estimation but the examples on their use concentrate on situations where the effect of a binary treatment is to be estimated.  
  
This paper aims to narrow the gap between the theory of causal models and practical data analysis by presenting an easy-to-follow example on the estimation of nonlinear causal relationships. As a starting point it is assumed that only the causal structure can be specified, i.e. the causal effects can be described qualitatively in a nonparametric form but not in a parametric form. Causal relationships are expected to have a complex nonlinear form which cannot be a priori modeled but is assumed to be continuous, differentiable and sufficiently smooth. The key idea is to first use causal calculus to find out the observational probabilities needed to calculate the causal effects and then estimate these observational probabilities from the data using flexible models from the arsenal of modern data analysis. The estimated causal effects cannot be expressed in a closed-form but can be numerically calculated for any input. The approach works if the required observational probabilities can be reliably interpolated or extrapolated from the model fitted to the data. Data missing at random creates an additional challenge for the estimation. Multiple imputation \citep{Rubin:multiple} of missing data works well with the idea of flexible nonparametric modeling because in multiple imputation an incomplete dataset is transformed to multiple complete datasets for which the parameters can estimated separately from other imputed datasets.

The estimation procedure is given in detail in Section~\ref{sec:procedure}. An example demonstrating the estimation of complex nonlinear causal effects is presented in Section~\ref{sec:examples}. Conclusions are given in Section~\ref{sec:conclusions}.

\section{Estimation of causal effects} \label{sec:procedure}

A straightforward approach for the estimation of causal effects is to first to define the causal model explicitly in a parametric form up to some unknown parameters and then use data to estimate these parameters. For instance, in linear Gaussian models, the causal effects are specified by the linear  models with unknown regression coefficients that can be estimated on the basis of the sample covariance matrix calculated from the data. This approach may turn out to be difficult to implement when the functional forms of the causal relationships are complex or unknown because it is not clear how the parametric model should specified and how its parameters could be estimated from the data. 

The estimation procedure proposed here utilizes nonparametric modeling of causal effects. 
It is assumed that the causal structure is known, the causal effects are complex and there are data collected by simple random sampling from the underlying population. An central concept in structural causal models is the do-operator \citep{Pearl:1995a,Pearl:book}. The notation $p(Y \mid \doo(X=x))$, or shortly $p(Y \mid \doo(x))$, refers to the distribution of random variable $Y$ when an action or intervention that sets the value of variable $X$ to $x$ is applied. The rules of causal calculus, also known as do-calculus, give graph theoretical conditions when it is allowed to insert or delete observations, to exchange actions and observation and to insert or delete actions. These rules and the standard probability calculus are sequentially applied to present a causal effect in the terms of observational probabilities.

The estimation procedure can be presented as follows:
\begin{enumerate}
 \item Specify the causal structure.
 \item Expand the causal structure to describe also the study design and the missing data mechanism.
 \item Check that the causal effect(s) of interest can be identified.
 \item Apply causal calculus to present the causal effect(s) using observational distributions.
 \item Apply multiple imputation to deal with missing data.
 \item Fit flexible models for the observational distributions needed to calculate the causal effect.
 \item Combine the fitted observational models and use them to predict the causal effect. 
\end{enumerate}
Step 1 is easy to carry out because the causal structure is assumed to be known. In Step~2, causal models with design \citep{dagdesign} can be utilized. In Steps~3 and 4, the rules of causal calculus are applied either intuitively or systematically via the identification algorithms \citep{tian2002general,shpitser2006identificationconditional,shpitser2006identificationjoint,Bareinboim:zidentifiability}. In Step~4, multiple imputation is applied in the standard way. In Step~5, fitting a model to the data is a statistical problem and data analysis methods in statistics and machine learning are available for this task. Naturally, validation is needed to avoid over-fitting. The step is repeated for all imputed data sets. In Step 6, the fitted observational models are first combined as the result derived in Step 3 indicates. This usually requires integration over some variables, which can be often carried out by summation with the actual data as demonstrated in the examples in Section~\ref{sec:examples}. Finally, the estimates are combined over the imputed datasets.


\section{Example: frontdoor adjustment for incomplete data} \label{sec:examples}
The example is chosen to illustrate challenges of causal estimation. Simulated data are used because this allows comparisons with the true causal model. The causal effects are set to be strongly nonlinear, which means that they cannot be modeled with linear models. The effects of the latent variables are set so that the conditional distributions in observational data strongly differ from the true uncounfounded causal effect. The setup is further complicated by missing data.

\begin{figure}
\begin{center}
\includegraphics[width=0.5\columnwidth]{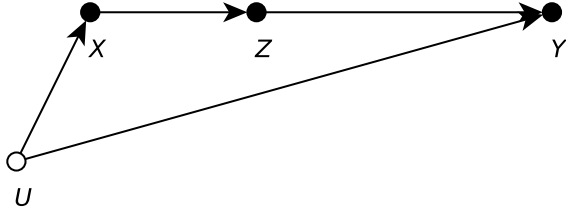}
\caption{Causal structure for the example. The solid circle indicates that the variable is observed and the open circle indicates that the variable is unobserved.  \label{fig:frontdoor}}
\end{center}
\end{figure}

The aim is to estimate the average causal effect $\E(Y \mid \doo(X=x))$ from data generated with the following structural equations
\begin{align*}
& U \sim \N(0,1), \\
& X' \sim \textrm{Unif}(-2,2), \\
& X = X'+U, \\
& Z = 4 \phi(X) + \epsilon_Z, \\
& Y = \phi(Z-0.5)+0.3*Z-0.1 U, \\
& \epsilon_Z \sim \N(0,0.1^2),
\end{align*}
where the nonlinear structural relationships are defined using the density function of the standard normal distribution
\begin{equation}
\phi(x)=\frac{1}{\sqrt{2 \pi}} \exp\left( -\frac{x^2}{2} \right).
\end{equation}
The corresponding causal structure is presented in Figure~\ref{fig:frontdoor}. The causal effect of $X$ to $Y$ is mediated through variable $Z$. In addition, there is an unknown confounder $U$ which has a causal effect both to $X$ and $Y$ but not to $Z$. Variables $X'$ and $\epsilon_Z$ in the structural equations are included only because of the convenience of the notation and are not presented in the graph.  \citet{Pearl:book} considers the same causal structure with an interpretation that $X$ represents smoking, $Z$ represents the tar deposits in the lungs and $Y$ represents the risk of lung cancer. Differently from Pearl, it is assumed here that all variables are continuous.

The observed data are denoted by variables $X^*$, $Z^*$ and $Y^*$. Variables $X^*$ and $Z^*$ are incompletely observed, which can be formally defined as 
\begin{align*}
 & X^*=
\begin{cases}
X, & \mbox{if } M_X=1 \\
\textrm{NA}, & \mbox{if } M_X=0,
\end{cases} \\
& Z^*=
\begin{cases}
Z, & \mbox{if } M_Z=1 \\
\textrm{NA}, & \mbox{if } M_Z=0,
\end{cases}
\end{align*}
where $M_X$ and $M_Z$ are missingness indicators. Response $Y$ is completely observed for the sample. The indicators $M_X$ and $M_Z$ are binary random variables that depend on other variables as follows
\begin{align}
& P(M_X=1 \mid \doo(Y=y)) = \Phi(2-y) \label{eq:M_X} \\
& P(M_Z=1 \mid \doo(Y=y)) = \Phi(4 y-1), \label{eq:M_Z}
\end{align}
where $\Phi$ stands for the cumulative distribution function of the standard normal distribution. The value of $X^*$ is more likely to be missing when $Y$ is large and the value of $Z^*$ is more likely to be missing when $Y$ is small.

\begin{figure}
\begin{center}
\includegraphics[width=\columnwidth]{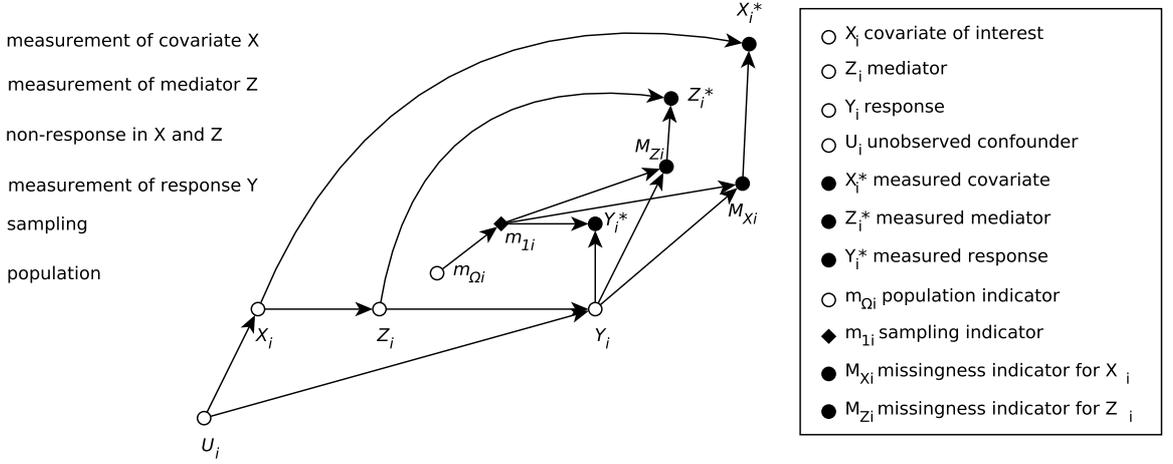}
\caption{Causal model with design for the example studied.  \label{fig:frontdoor_with_design}}
\end{center}
\end{figure}

A causal model with design that combines the causal model and the missing data mechanism is presented in Figure~\ref{fig:frontdoor_with_design}. The causal model is the same as in the graph in Figure~\ref{fig:frontdoor}. Subscript $i$ indexes the individuals in a finite well-defined closed population $\Omega=\{1,\ldots,N\}$. Variable $m_{\Omega i}$ represents an indicator for the population membership and is defined as $m_{\Omega i}=1$ if $i \in \Omega$ and $m_{\Omega i}=0$ if $i \notin \Omega$. The data are collected for a sample from population $\Omega$. Indicator variable $m_{1i}$ has value 1 if the individual is selected to the sample and 0 otherwise. Variables $X^*$, $Z^*$ and $Y^*$ are recorded and are therefore marked with a solid circle. The underlying variables in the population, $X_i$, $Z_i$ and $Y_i$ are not observed are therefore marked with an open circle.  The value of response $Y_i^*$ is observed for the whole sample. The value of covariate $X_i^*$ is observed when $M_{Xi}=1$ where $M_{Xi}$ depends on $Y_i$ as indicated in equation~\eref{eq:M_X}. The value of response $Z_i^*$ is observed when $M_{Zi}=1$ where $M_{Zi}$ depends on $Y_i$ as indicated in equation~\eref{eq:M_Z}.  

\begin{figure}
\begin{center}
\includegraphics[width=0.9\columnwidth]{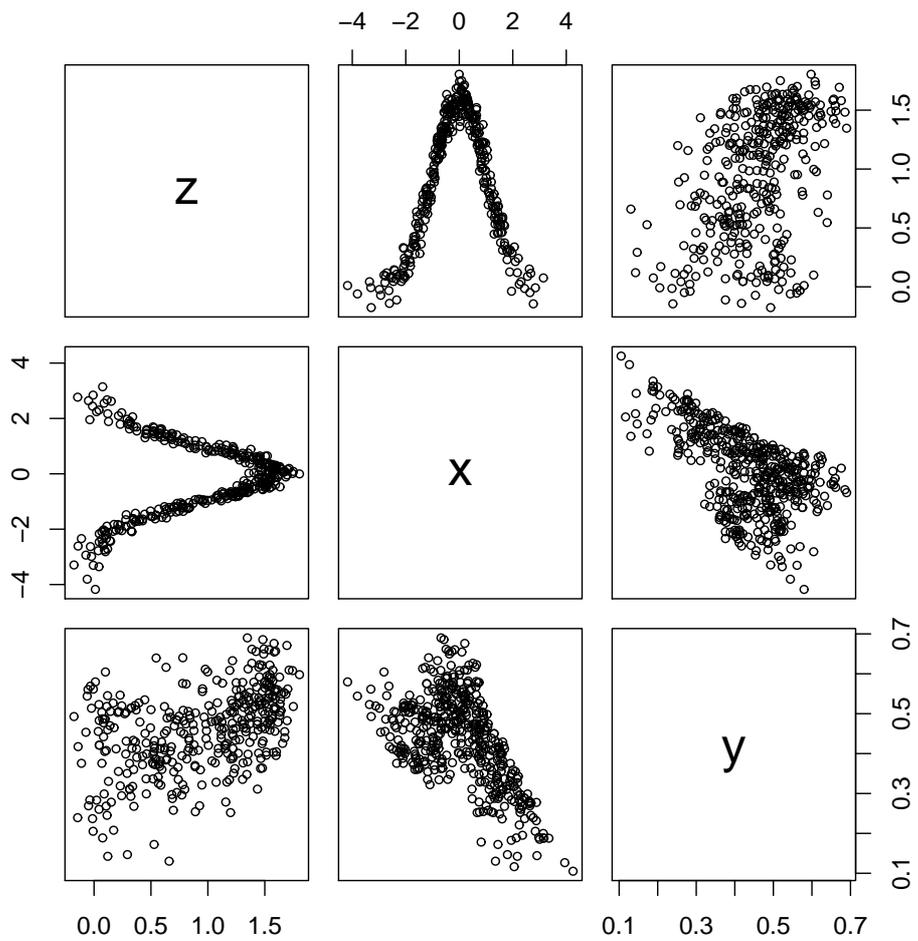}
\caption{Scatterplot matrix of a subsample of 500 observations.  \label{fig:frontdoor_scatterplot}}
\end{center}
\end{figure}

\begin{figure}
\begin{center}
\includegraphics[width=0.8\columnwidth]{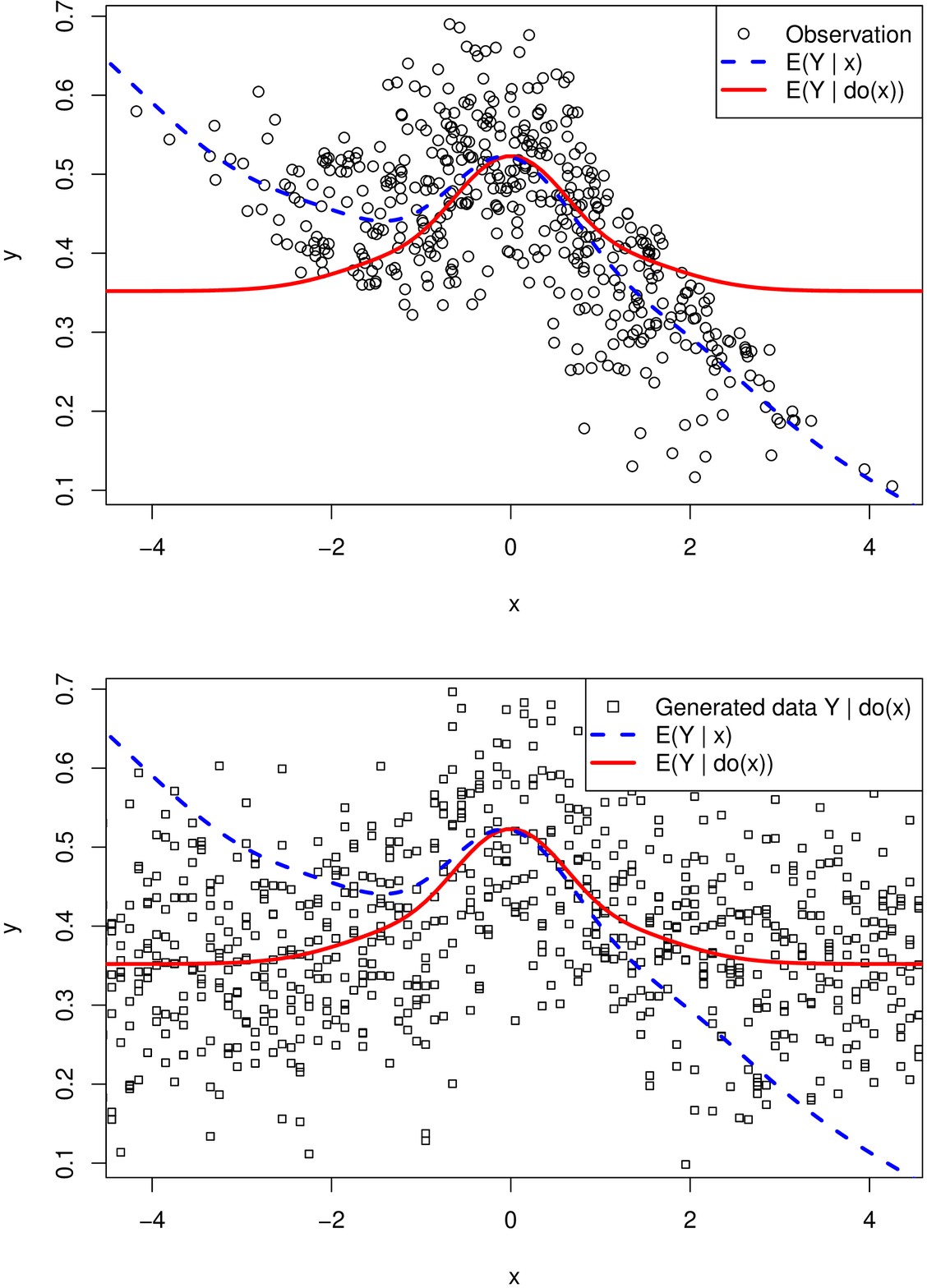}
\caption{Average causal effect compared to average conditional mean.  In the upper panel, the true average causal effect (solid line) is presented together with observations (subsample of 500 observations) and their conditional average (dotted line). In the lower panel, the same curves are presented with data generated from the hypothetical experiment where the values of $X$ are controlled and eight values of response $Y$ are generated for each $x$.  \label{fig:data_and_true_causal}}
\end{center}
\end{figure}

The input available for the estimation contains only the graph in Figure~\ref{fig:frontdoor_with_design} and the data $(X_i^*,Z_i^*,Y_i^*)$, $i=1,\ldots,n=20000$. In the data, the value of $X_i^*$ is missing in 6\% of cases, the value of $Z_i^*$ is missing in 26\% of cases and both $X_i^*$ and  $Z_i^*$ are missing in 1.5\% of cases.   The joint distribution of  $Z^*$, $X^*$ and $Y^*$ is illustrated in scatterplots in Figure~\ref{fig:frontdoor_scatterplot}. The most discernible feature of the joint distribution is the strong nonlinear association between covariate $X^*$ and mediator $Z^*$. Both $X^*$ and $Z^*$ are also associated with response $Y^*$. Figure~\ref{fig:data_and_true_causal} shows the true causal effect of interest together with the observational data. From the observational distribution $p(Y \mid x)$, one might misjudge that $Y$ decreases as the function $X$ but in reality the observed decreasing trend is due to the unobserved confounder $U$. A particular challenge is that certain combinations of the variable values are not present in the data. For instance, if the value of $X$ is set to 3, the average response is around 0.35 but in the data there no observations even close to this point.

It is possible to identify average causal effect $\E(Y \mid \doo(X=x))$ from observational data because in the causal structure there exists no bi-directed path (i.e. a path where each edge starts from or ends to a latent node) between $X$ and any of its children \citep{tian2002general,Pearl:book}. Here, $Z$ is the only child of $X$ and there are no edges between $Z$ and $U$. Applying the rules of causal calculus \citep{Pearl:book}, the causal effect of $X$ to $Y$ can be expressed as
\begin{equation} \label{eq:frontdoor}
p(Y \mid \doo(X=x))= \int p(z \mid X=x) \int p(Y \mid X=x',Z=z) p(X=x') \dd x' \dd z
\end{equation}
and the average causal effect is obtained as
\begin{equation} \label{eq:frontdoorE}
\E(Y \mid \doo(X=x))= \int p(z \mid X=x) \int \E(Y \mid X=x',Z=z) p(X=x') \dd x' \dd z.
\end{equation}
This result is known as the frontdoor adjustment \citep{Pearl:1995a}. The derivation is presented in detail in \citep[p. 81--82]{Pearl:book}. Equation~\eref{eq:frontdoor} means that the problem of estimating causal effect $\E(Y \mid \doo(X=x))$ can be reduced to the problem of estimating observational distributions $p(z \mid X=x)$, $p(Y \mid X=x,Z=z)$ and $p(X=x')$.

Before proceeding with the estimation the problem of missing data must be addressed. Complete case analysis is expected to lead biased results because the missingness of $X$ and $Z$ depends on response $Y$. Using the d-separation criterion \citep[p. 16--17]{Pearl:book}, it can be concluded from the causal model with design in Figure~\ref{fig:frontdoor_with_design} that on the condition of $Y$ missingness is independent from the missing value
\begin{align*}
&  M_{X} \independent X \mid Y \\
&  M_{Z} \independent Z \mid Y.
\end{align*}
As variable $Y$ is fully observed in the sample, the data are missing at random and multiple imputation can be applied. Multiple imputation is not the only way to approach the missing data problem but it is easy to implement together with nonparametric estimation of causal effects. Due to nonlinearity of the dependencies, special attention has to be paid on the fit of the imputation model. The MICE-algorithm \citep{vanbuuren2012flexible} implemented in the R package \texttt{mice} \citep{vanbuuren:mice} is applied in the imputation. The default method of the package, predictive mean matching with linear prediction models, fails because of the for each value of $Z$, the conditional distribution of $X$ is bimodal. As a solution, the imputation model uses $|X|$ and $\textrm{sign}(X)$ instead of $X$ and generalized additive models (GAM) \citep{hastie1990generalized} are applied as prediction model in predictive mean matching. The fit of the imputations is checked by visually comparing the observed and imputed values.

The frontdoor adjustment~\eref{eq:frontdoor} leads to estimators
\begin{align} \label{eq:frontdoor_estimator}
& \hat{p}(Y \mid \doo(X=x)) = \frac{1}{n} \sum_{i=1}^n \hat{p} \left( Y \mid X=x_i,Z=\tilde{z}_i \right) \\
& \hat{\E}(Y \mid \doo(X=x)) = \frac{1}{n} \sum_{i=1}^n \hat{\E} \left( Y \mid X=x_i,Z=\tilde{z}_i \right)
\end{align}
where $\tilde{z}_i$ is a random variable generated from the estimated distribution $\hat{p}(Z \mid X=x)$. The distributions $\hat{p}(Z \mid X=x)$ and $\hat{p} \left( Y \mid X=x,Z=z \right)$ are modeled by GAM with smoothing splines \citep{Wood:GAMbook}. The estimation is carried out using the R package \texttt{mgcv} \citep{Rmgcv} which includes automatic smoothness selection. The estimation is repeated for ten imputed datasets and a complete case analysis is carried out as a benchmark. The random variables from the estimated distributions are generated by adding resampled residuals to the estimated expected value. The estimated models have a good fit with the observational data.

\begin{figure}
\begin{center}
\includegraphics[width=0.8\columnwidth]{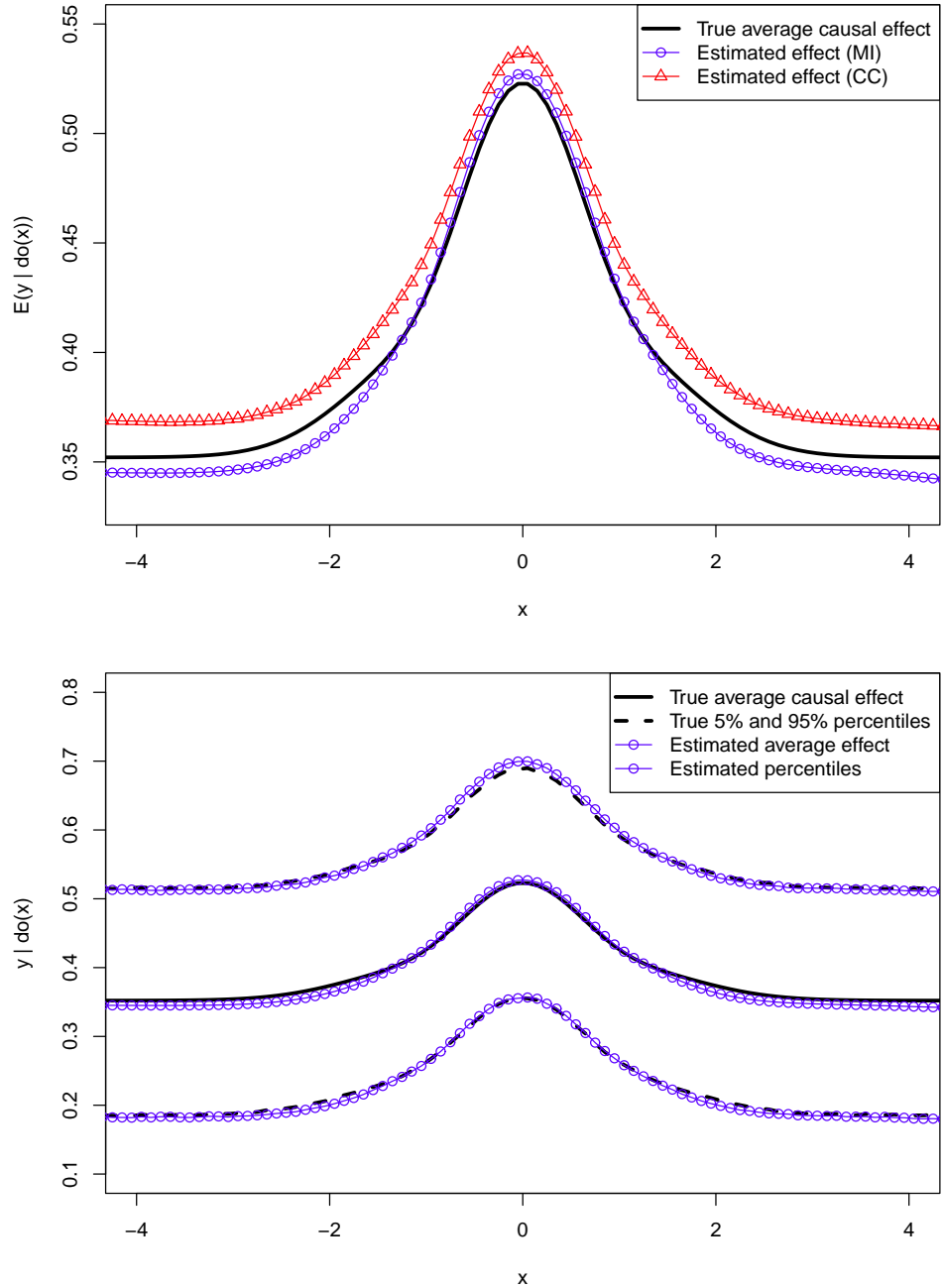}
\caption{Estimated causal effects compared to the true causal effect. The upper panel shows the average causal effect estimated using multiple imputed data (MI) or only completely cases (CC). The lower panel shows the true and the estimated 5\% and 95\% percentiles of the causal distribution $p(Y \mid \doo(X=x))$ . \label{fig:frontdoor_results}}
\end{center}
\end{figure}

The estimated causal effects are presented in Figure~\ref{fig:frontdoor_results}. Both the average causal effect $\E(Y \mid \doo(X=x))$ and the causal distribution $p(Y \mid \doo(X=x))$ were estimated without a significant bias when $x \in (-2,2)$ and multiple imputation was applied. Complete case analysis systematically over-estimated the average causal effect, which demonstrates the importance of proper handling of missing data. Due to the relatively large size of data, the sample variability was not an issue here. It can be concluded that the chosen nonparametric strategy was successful in the estimation of the average causal effects.

\section{Conclusions} \label{sec:conclusions}
It was shown how causal effects can estimated by using causal models with design, causal calculus, multiple imputation and generalized additive models. 
It was demonstrated that the estimation of causal effects does not necessarily require the causal model to be specified parametrically but it suffices to model the observational probability distributions. The causal structure can be often specified even if it is very difficult to forge a suitable parametric model for the causal effects. The data can modeled with flexible models for which ready-made software implementations exist.
The key message can be summarized in the form of the equation
\begin{equation*}
 \textrm{Causal estimation} = \textrm{causal calculus} + \textrm{data analysis}.
\end{equation*}

The applied nonparametric approach requires that the causal effects are sufficiently smooth to be estimated from the data. In the example, the variables were continuous but the approach works similarly for discrete variables. Large datasets may be required if the causal effects are complicated, the noise level is high or there are unobserved confounders. If the data are missing not at random, external information on the missing data mechanism is needed. The interpretation of the causal estimates requires care. The results should be evaluated with respect to the existing scientific knowledge before any conclusions can be made. 



The presented example hopefully encourages statisticians to target causal problems. Tools for causal estimation exist and are not too difficult to use.


\bibliographystyle{apalike}
\bibliography{dag,omat}

\end{document}